\input{aipcheck}

\documentclass[
    ,final            
  ]
  {aipproc}
  
  \usepackage{amsmath,amssymb}
  \usepackage{floatflt}

\layoutstyle{6x9}

\begin{document}

\title{All-loop calculations of total, elastic and single
diffractive cross sections in RFT via the stochastic
approach.}

\classification{ 12.40.Nn, 
13.85.Dz, 
13.85.Hd, 
13.85.Lg 
}
\keywords      {Reggeon Field theory, Pomeron loops, reaction-diffusion methods, diffractive cross sections}

\author{R.S.Kolevatov$^1$\setcounter{footnote}{1}\footnotetext{Also at Saint-Petersburg State University
  (Ulyanovskaya 1, 198504, Saint-Petersburg, Russia)
  and University of Oslo (PB1048 Blindern, N-0316 Oslo, Norway).
  }}{
  address={SUBATECH, Ecole des Mines de Nantes, 4 rue Alfred Kastler, 44307 Nantes Cedex 3, France}
}

\author{K.G.Boreskov}{
  address={Institute of Theoretical and Experimental Physics, 117259, Moscow, Russia}
}

\def\d{\partial}
\def\b{{\bf b}}
\def\k{{\bf k}}
\def\z{{\bf z}}
\def\q{{\bf q}}
\def\u{{\bf u}}
\def\B{{\cal B}}
\def\D{{\cal D}}
\def\K{{\cal K}}
\def\Z{{\cal Z}}
\def\X{{\cal X}}
\def\Y{{\cal Y}}
\def\Q{{\cal Q}}
\def\II{{\cal I}}
\def\F{{\cal F}}
\def\G{{\cal G}}
\def\y{\tilde{y}}
\def\h{\tilde{h}}
\def\f{\tilde{f}}
\def\w{\tilde{w}}
\def\g{\tilde{g}}
\def\s{\tilde{s}}
\def\tb{\tilde{\bf b}}
\def\T{\hat{T}}
\def\H{\hat{H}}
\def\p+{\! + \!}
\def\m-{\! - \!}
\def\r={\! = \!}
\def\Im{{\mathrm{Im}\,}}
\def\Re{{\mathrm{Re}\,}}
\def\ds{\displaystyle}
\def\scs{\scriptstyle}
\def\scss{\scriptscriptstyle}
\def\N{{\scss N}}
\def\A{{\scss A}}
\def\phid{\phi^\dagger}
\def\r3P{r_{{\scss 3P}}}
\def\Aph{A\phantom{\widetilde{A}\!\!\!\!}}
\def\GeV{~\mathrm{GeV}}
\def\fm{~\mathrm{fm}}

\begin{abstract}
We apply the stochastic approach to the calculation of the Reggeon Field Theory (RFT) elastic amplitude and its single diffractive cut. The results for the total, elastic and single difractive cross sections with account of all Pomeron loops are obtained.
\end{abstract}

\maketitle


\section{The stochastic approach}

The elastic scattering amplitude of the RFT \cite{Gribov'64} is given in terms of quasiparticle exchanges (Pomerons and Reggeons), its cuts determine cross sections of various inelastic processes. While the lower energy data on total and elastic cross sections can be described in terms of non-interacting Pomerons and Reggeons, the high energy behaviour of total, elastic and  high-mass diffractive cross sections 
makes the account of Pomeron interactions and loops in the RFT fits absolutely essential.
For this task we apply a method which is based on the approach known as reaction-diffusion (RD) or stochastic. 

It was observed \cite{Grassberger:1978pr} that a stochastic system of classical particles in the 2-dimensional plane admits a field-theoretical description with the Lagrangian of the Gribov--Regge theory with triple and $2\to2$ interaction terms:
\begin{equation}
 \mathcal{L} = \frac{1}{2} \phi^\dagger (\overleftarrow{\partial_y} - \overrightarrow{\partial_y})\phi
 - \alpha' (\nabla_\b \phid)(\nabla_\b \phi)   + \Delta \phid\phi + i\,\r3P \phi^\dagger \phi (\phi^\dagger + \phi) + \chi {\phi^\dagger}^2 \phi^2 
\label{eq:LRFT}
\end{equation}
The classical partons are allowed to move chaotically (characterized by diffusion coefficient $D$), split, $A\rightarrow A+A$, with a splitting probability per unit time $\lambda$, or die, $A\rightarrow \emptyset$, with a death probability $m_1$. When two partons are brought within the reaction range $a$ due to the diffusion, they can pairwise fuse, $A+A\rightarrow A$, or annihilate, $A+A\rightarrow \emptyset$ with the rates $\nu$ and $m_2$ correspondingly. A set of inclusive $s$-parton distributions:
\begin{equation}
 f_s(y;\Z_s)= \sum_N \cfrac{1}{(N-s)!} \int\! d\B_N \;\rho_\N(y;\B_N)
 \prod_{i=1}^{s} \delta(\z_i-\b_i), \label{def_f}
\end{equation}
obeys the same evolution equations as the set of the exact Green functions for the Lagrangian \eqref{eq:LRFT}. Here  $\rho_\N(y;\B_N)$ are the symmetrized probability densities with normalization 
$
\sum_N \frac{1}{N!}\int d\B_\N \rho_\N (y;\B_\N) = \sum_M p_\N (y) =1
$  and $\Z_s\equiv\{\z_1,\ldots, \z_s\}$. Evolution time for partonic system in RD approach is dual to the rapidity in the RFT.

Phenomenological parameters of the Lagrangian \eqref{eq:LRFT} have direct correspondence with the rates of the stochastic system (see table 1). The parton interaction distance $a$ serves as a regularization parameter for the Pomeron loops. For given values of the coupling $r_{3P}$ and the scale  $\epsilon \equiv \pi a^2$ the quartic coupling $\chi$ can be varied. 
\begin{table}[h]
\centering
\caption{Relation between the parameters of the RFT and those of the stochastic approach.}
\begin{tabular}{|c||c|c|c|c|c|c|}
 \hline
 RFT & $\alpha'$ & $\Delta$ & $\r3P$, $\mathbb P$ splitting vertex& $\r3P$, $\mathbb P$ fusion vertex & $\chi$, $2{\mathbb P}\to 2{\mathbb P}$\\
 \hline
 RD-approach & $D$ & $\lambda - m_1$ & $\lambda \sqrt{\epsilon}$ & $({m_2}+\tfrac{1}{2}{\nu})\sqrt{\epsilon}$ &  $\tfrac{1}{2} ({m_2}+{\nu})\epsilon $\\
 \hline
\end{tabular}
\label{TAB:RFT-ST-rel}
\end{table}
\vskip-1mm
This equivalence allows to compute the exact Green functions of the RFT with the account of all loop contributions following a Monte-Carlo evolution of the RD system. 


\section{Amplitude calculation}

The paper \cite{Boreskov:2001nw} suggested a straightforward approach to the amplitude calculation. One has to compute the $n$-point Green functions in convolution with hadron--$n$-Pomeron vertices (grey blocks in fig. 1{\it a}) for the projectile ($f_s$) and target($\tilde f_s$) and perform their numerical convolution according to the general rules of the Reggeon field theory at some linkage point $y$ in rapidity.  In particular, the elastic scattering amplitude is : \vskip-5mm
\begin{equation}
 T^{\rm el}({\bf b}, Y) 
 =\sum_{s=1}^\infty \frac{(-1)^{s-1}} {s!} \int d\Z_s
 d\tilde \Z_s f_s(y;\Z_s ) \tilde f_s(Y-y;\tilde \Z_s) \prod_{i=1}^s g ({\bf z}_i - \tilde {\bf z}_i - {\bf b}) .
\label{TST}
\end{equation} 
\vskip-2mm
\noindent Here $g$ are some narrow functions normalized to $\int g({\bf b}) d^2{\bf b}=\epsilon$.

The most efficient way to compute the amplitude is its computation on the event by event basis with subsequent Monte-Carlo average setting the linkage point to the target rapidity. The inclusive $n$-parton distributions at projectile and target rapidities coincide with the hadron--$n$-Pomeron vertices \cite{Kolevatov:2011nf}:
\vskip-4mm
\begin{equation}
 f_s(y=0;\Z_s)  
\equiv \mu_s p_s(\Z_s)=  \epsilon^{s/2} \mathcal{N}^{(s)}(\Z_s) , \label{fs-vertex}
\end{equation}
This gives the distribution of partons at zero evolution time in number and positions in the transverse plane. In accordance with it the initial configurations of partons should be generated within the first step of the Monte-Carlo averaging procedure. 

Upon the Monte-Carlo evolution of initial random parton configuration one gets a set of $N$ partons at certain positions $\hat b_i$ in the transverse plane. The event realization of inclusive distribution is thus  
$ f^{event}_s (\B_s)= \sum_{\{i_1\ldots i_{s}\}\in \{ 1\ldots N\}} \delta(b_1-\hat b_{i_1})\ldots \delta (b_{s} - \hat b_{i_{s}})$ and upon convolution with the set of target--$n$-Pomeron vertexes leads to: \vskip-4mm
\begin{equation}
T_{\rm sample}^{\rm el} ({\bf b}) = \sum_{s=1}^{N} (-1)^{s-1} \tilde \mu_s \epsilon^s \sum_{i_1<i_2 \ldots <
i_s} \tilde p_s(\hat {\bf x}_{i_1} - {\bf b}, \ldots ,\hat {\bf x}_{i_s}-{\bf b}). \label{T-onesample}
\end{equation} \nopagebreak The actual value of the elastic amplitude as a function of the impact parameter $b$ is computed by making Monte-Carlo average of \eqref{T-onesample}.

\setlength{\textfloatsep}{15pt plus 2pt minus 4pt}
\setlength{\floatsep}{15pt plus 2pt minus 4pt}

\begin{figure}[htbp]
 \centering
 \includegraphics[width=0.8\hsize]{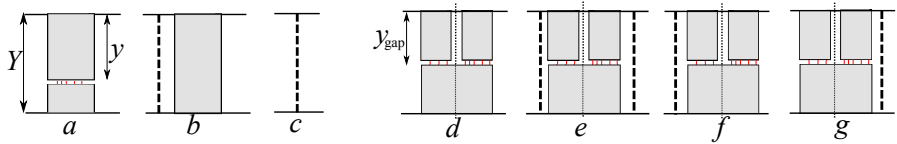}
 \caption{Diagrams taken into account in calculation of elastic amplitude ({\it a--c}) and its SD cut ({\it d--g}).}
\end{figure}


A single diffractive cut of the amplitude for a given value of the rapidity gap $y_{\rm gap}$ (with elastic contribution included) can be computed as a sum of two terms \footnote{We are currently preparing a paper with a complete derivation of eq. \eqref{eq:SDcut}.}:
\begin{equation}
T^{\rm SD cut} ({\bf b},Y,y_{\rm gap})= 2 T^{\rm el}({\bf b},Y) - T'({\bf b},Y,y_{\rm gap}). \label{eq:SDcut}
\end{equation}
The term $T'({\bf b},Y,y_{\rm gap})$ is computed in the same way as the elastic amplitude by making a Monte-Carlo average of \eqref{T-onesample} with only the distinction in preparation of the projectile-associated set of partons. The evolution starts with \emph{two sets} which evolve \emph{independently} up to the evolution time $y_{\rm gap}$ corresponding to the width of the rapidity gap. At that point the resulting partons are \emph{combined into a single set} which further evolves in the standard way from $y_{\rm gap}$ up to the target rapidity $Y$.

\section{Parameters and results}
We use the approach described above to compute the elastic amplitude in impact parameter representation and its single diffractive cut with the account of Pomeron trajectories only. For the parameterization of the proton--$n$-Pomeron vertexes we use the two-channel eikonal approximation with the same values of the channel weights $C_1=C_2=0.5$ and of the parameter $\eta=0.55$ for the relation between the channel couplings to the Pomeron $\beta_{1/2}=(1\pm\eta)\beta_0$ as in \cite{Martin+Luna'09}. We however use a Gaussian parametrization of these vertexes and neglect the real part of the Pomeron exchange amplitude, $\Im A_P(b) = T^{\rm el}(b)$. The triple coupling value $r_{3P}=0.087$~GeV$^{-1}$ is taken according to \cite{Kaidalov:1979jz}\footnote{Our normalization of $r_{3{\mathbb P}}$ (see \cite{Kolevatov:2011nf}) differs from that in \cite{Martin+Luna'09} and \cite{Kaidalov:1979jz}.}. 

\begin{floatingtable}{
\begin{tabular}{|l|ccc|}
 \hline Trajectory & $\mathbb{P}$ & $R_+$ & $R_-$ \\
\hline  $\Delta_{0/+/-}$ &  0.195 & -0.34  &   -0.55 \\ 
  $\alpha'$, GeV$^{-2}$ &  0.154 & 0.70 &  1.0\\
  $R^2$, GeV$^{-2}$ & 3.62 & 3.0 &   5.2 \\
  $\beta_{0/+/-}$, GeV$^{-1}$ & 4.7 & 4.05 &   2.59 \\
  \hline
\end{tabular}}
\noindent \footnotesize {\bf TABLE 2.} Parameters of the trajectories and their couplings to proton.
\end{floatingtable}In order to obtain a better description of the data at lower energies we add a lowest order contribution of secondary trajectories of positive and negative signature (see fig. 1). This contribution is added numerically to the all-loop Pomeron exchange amplitude given by the Monte-Carlo computation as described above making use of the Regge factorization. We also keep the real part for the secondary Reggeon contribution.


We perform a fit of total and elastic cross sections together with elastic scattering slope. For this we in addition fix apriori the regularization scale $a=0.036$~fm$=0.182$~GeV$^{-1}$ and $2\to 2$ coupling $\chi$ by setting $\nu=1/2\lambda$ and $m_2=0$ (see tab.~1).  This gives $\chi =2.87\times 10^{-2}$~GeV$^{-2}$. This set is referred to as ``set~3'' in fig.~2. The parameters of Regge trajectories used and couplings to proton are listed in tab.~2. 

To illustrate the dependence of the result on the regularization scale and the quartic coupling we perform calculation also for the other parameter sets which differ from the original (set 3) by values of $a$ and $\chi$ and have the same parameters of the Regge trajectories (sets 1, 2, 4). Namely we take $a_1=a_2=0.091$~GeV$^{-1}$, $a_4=0.182$~GeV$^{-1}$; $\chi_1=\chi_4  = 1.435\times 10^{-2}$~GeV$^{-2}$,  $\chi_2 = 7.17\times 10^{-3}$~GeV$^{-2}$. 

In fig.~3 (left plot) we show results of the diffractive cross sections calculation for the sets 1--4 (for the accounted graphs see fig.\,1\,{\it d--g}). Plots for the inelastic and diffractive\linebreak profiles indicate that at high energies the inelastic diffractive contribution comes dominantly from the ring on the periphery of the disc. This is consistent with the logarithmic growth of the all-loop inelastic diffractive cross sections with collision energy.
\vskip-4mm



\begin{figure}
\centering
\includegraphics[width=\hsize]{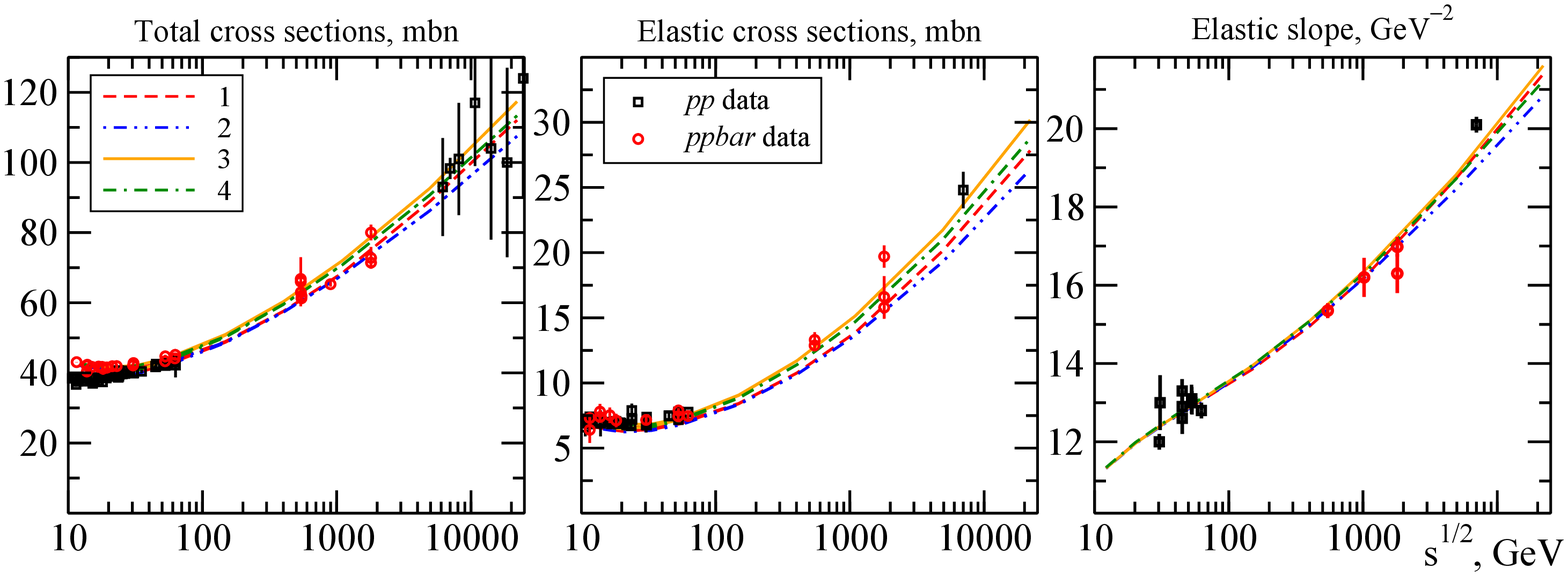}
\caption{Total cross section, elastic cross section and elastic scattering slope.}
\end{figure}
\begin{figure}
\centering
\includegraphics[width=\hsize]{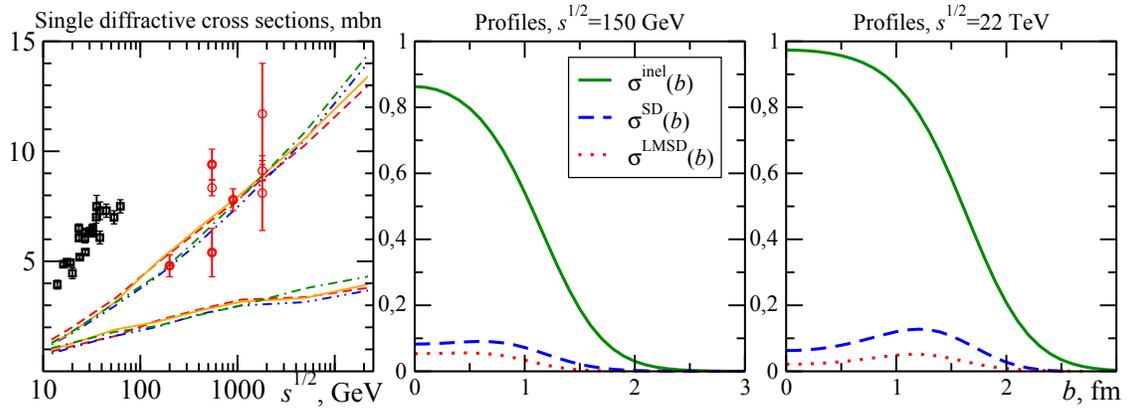}
\caption{{\it Left}: single diffractive cross section for $M^2<0.05s$ ($y_{\rm gap}=3$, upper curves) and low-mass ($y_{\rm gap}=2\ln \sqrt{s}$, lower curves) dissociation. {\it Center, right}: unintegrated profiles for inelastic and single diffractive cross sections, high- and low-mass (for set 3).}
\end{figure}



\begin{theacknowledgments}
Authors are thankful to O.\,V.\,Kancheli and L.\,V.\,Bravina. Work of RK was supported by the NFR Project 185664/V30 and the RFBR grant 12-02-00356-a.
\end{theacknowledgments}



\bibliographystyle{aipproc}   


\begin{thebibliography}{9}

 \bibitem{Gribov'64} V. N. Gribov,
 \emph{Sov. Phys. JETP} \textbf{26}, 414 (1968) [\emph{Zh. Eksp. Teor. Fiz.} \textbf{53}, 654 (1967)].
\bibitem{Grassberger:1978pr}
  P.~Grassberger, K.~Sundermeyer,
  \emph{Phys.\ Lett.} {\bf B77 } (1978)  220.
\bibitem{Boreskov:2001nw}
  K.~G.~Boreskov,
  In Olshanetsky, M. (ed.) et al.: \emph{Multiple facets of quantization and supersymmetry}, pp.~322--351.
  [hep-ph/0112325].

\bibitem{Kolevatov:2011nf} 
  R.\,S.\,Kolevatov, K.\,G.\,Boreskov and L.\,V.\,Bravina,
  \emph{Eur.\ Phys.\ J.\ C} {\bf 71}, 1757 (2011).
  [arXiv:1105.3673 [hep-ph]].

    
\bibitem{Martin+Luna'09}
 E.G.S. Luna, V.A. Khoze, A.D. Martin and M.G. Ryskin,
 \emph{Eur. Phys. J. C} \textbf{59} (2009), pp.~1--12.
  [arXiv:0807.4115 [hep-ph]].
  
  \bibitem{Kaidalov:1979jz}
  A.~B.~Kaidalov,
  \emph{Phys.\ Rept.} {\bf 50 }, pp.~157--226 (1979).
  

\end{thebibliography}



\end{document}